# The negative effects of citing with a national orientation in terms of recognition: national and international citations in natural-sciences papers from Germany, the Netherlands, and the UK

Lutz Bornmann*, Jonathan Adams**, & Loet Leydesdorff***

*Corresponding author:
Division for Science and Innovation Studies
Administrative Headquarters of the Max Planck Society
Hofgartenstr. 8,
80539 Munich, Germany.
Email: bornmann@gv.mpg.de

** The Policy Institute at King's,
King's College London,
22 Kingsway,
London WC2B 6LE, UK.
(and)
Institute for Scientific Information, Clarivate Analytics
Email: jonathan.adams@clarivate.com

***Amsterdam School of Communication Research (ASCoR)
University of Amsterdam
PO Box 15793,
1001 NG Amsterdam, The Netherlands.
Email: loet@leydesdorff.net

**Abstract**

Nations can be distinguished in terms of whether domestic or international research is cited. We analyzed the research output in natural sciences of three leading European research economies (Germany, the Netherlands, and the UK) and ask where their researchers look for the knowledge that underpins their most highly-cited papers. Is one internationally oriented or is citation limited to national resources? Do the citation patterns reflect a growing differentiation between the domestic and international research enterprise? To evaluate change over time, we include natural-sciences papers published in the countries from three publication years: 2004, 2009, and 2014. The results show that articles co-authored by researchers from Germany or the Netherlands are less likely to be among the globally most highly-cited articles if they also cite "domestic" research (i.e. research authored by authors from the same country). To put this another way, less well-cited research is more likely to stand on domestic shoulders and research that becomes more highly-cited is more likely to stand on international shoulders. A possible reason for the results is that researchers "over-cite" the papers from their own country – lacking the focus on quality in citing. However, these differences between domestic and international shoulders are not visible for the UK.

# 1 Introduction

The sciences develop internationally, but the funding is mainly national. In a time of "America first" and similar developments in other countries, national governments are challenged to legitimate funding in terms of national priorities. The tensions and trade-offs between international and national perspectives can be expected to differ among disciplines. While one can legitimately dispute positivism in "German sociology" (Adorno et al., 1969; Leydesdorff & Milojević, 2015), alternative approaches in physics, e.g., "Deutsche Physik" or biology on the basis of national aspirations can be considered obscure (Graham, 1974; Lecourt, 1976). However, there can be a tension between national and international research agendas. Hagendijk and Smeenk (1989) used the metaphor of "national subfields" which may be specific in the dependency on domains and resources like a specific lake district. Merton (1973) distinguished between the development of the international literature and national "styles" in the social sciences responsive to local conditions. Scientific elites can play a mediating role in appeasing the tensions that emerge between national resources and international main-stream research (Mulkay, 1976).

Adams (2013) has argued that there is a "fourth age" of research in which the growing divide between international and domestic research will influence each nation's ability to draw on the global knowledge base and influence its national scientific wealth. From this perspective, one can expect that a comparative analysis of references in highly-cited papers may show some differences in the degree to which the most impactful (relatively highly-cited) research and its mainstream "platform" research might draw on an international or a relatively domestic knowledge base. Policy interventions might be deemed necessary where the disparity and connection between the domestic base and the international network grows too large and; the related management considerations might apply equally at institutional as at national levels.

In this study, we review a diversity of specific bibliometric studies at the country level and identify a gap of policy significance. We suggest that one needs to ask not only "which country produces the highly-cited papers" but also "can we determine the countries on whose research the highly-cited papers build"? Whom do researchers cite given the pressures to maintain both a national and international profile? Is the orientation *tout-court* international (Merton, 1942) or are national contributions nevertheless cited above expectation? Abramo and D'Angelo (2018) noted that country affiliations of the citing authors can be used to trace the countries benefiting from results produced in a national research system.

To test national benefits we focus on the research output of three leading European research economies in natural sciences and ask where their researchers look for the knowledge that underpins their most highly-cited papers. Is this restricted to national resources or does it reflect a growing differentiation between the domestic and the international research enterprise? Which implications does this have for growing international networks and the way knowledge is shared? And does the outcome indicate differences in the degree to which each country's knowledge is useful for itself and other countries?

## 2 Literature overview

Bibliometric results at the national or country level can be found not only in research papers (e.g. Bornmann, Wagner, & Leydesdorff, 2015), but also in reports (e.g. Kamalski et al., 2017; Michels, Fu, Neuhäusler, & Frietsch, 2014; National Science Board, 2016), in news items (e.g. Marshall & Travis, 2011; Van Noorden, 2014) and on web sites (see statistics e.g. by SCImago at e.g. http://www.scimagojr.com/countryrank.php or by *Nature* at https://www.natureindex.com). The foundation of most studies, published in print or on the Web, is a global comparison of national publication outputs and citation impacts.

An alternative focus for studies that do not address the global research system, may be (1) specific countries, such as China, the UK, and the USA, (2) specific alliances, such as Europe, and (3) specific country types, such as emerging economies. For example, Leydesdorff and Wagner (2009) and Wagner (2011) studied the dominance of the USA in the global science system; while Adams (2010) investigated the international comparative performance of the UK. Adams, Pendlebury, and Stembridge (2013) reported on the global research and innovation impact of the BRICK economies (Brazil, Russia, India, China and South Korea). Aagaard and Schneider (2016) analyzed the relationship between research policy inventions and academic performance in Denmark.

Many analyses concur with established views of relative national performance, but some lead to controversial conclusions. Rodríguez-Navarro and Narin (2018) address the so-called EU paradox of high scientific performance (in terms of bibliometric indicators) but apparently low innovation performance (in terms of technology indicators) (see also Rodriguez-Navarro & Brito, 2018). The authors suggest that the paradox rests on a false assumption based on erroneous performance indicators (i.e., the use of simple publication counts). The authors argue that it is just the frequently-cited papers that critically underpin innovations. On this indicator, the EU falls behind the USA in research performance.

Many country-level studies include China because of the disruptive effect of its economic growth on a previously stable world order. Most of these studies describe China's explosive increase in publications (e.g. Xie, Zhang, & Lai, 2014) whilst finding that citation impact remains relatively low (e.g. Leydesdorff, Wagner, & Bornmann, 2014; Wagner, Bornmann, & Leydesdorff, 2015; Zhou & Bornmann, 2015). However, Confraria, Godinho, and Wang (2017) use more recent data and find that "the average Chinese citation impact is very close to the world average, and that China is already performing considerably better than the world average in some scientific areas, such as 'Agricultural Sciences'; 'Engineering'; 'Mathematics'; 'Plant & Animal Science', and 'Social Sciences'" (p. 269). The reasons for China's rise in research performance have been discussed (Sun & Cao, 2014) and suggestions made around increasing the quality of research (Yang, 2016).

Many bibliometric studies at the country level use only simple indicators such as raw paper counts and citation averages. Some studies, however, have investigated bibliometric data in more elaborate and revealing ways, e.g. by using country-shares of world citations in relation to shares of publications. Hassan and Haddawy (2013) explored the knowledge flows among countries and developed the web-based tool Knowledge-MAPPER. Based on a new source of bibliometric data (Microsoft Academic), Dong, Ma, Shen, and Wang (2017) presented numbers on country-shares of global citations and related them to productivity: "During the early 20th century, the US, Germany, and the UK created 95% and collected 97% of the world's citations, while these two shares were decreased by about half as of the 21st century, to 46% and 58%, respectively".

The critical characteristic of all these, very diverse, studies is that they look at a country as an entity in a global set of similar entities. They do not consider interactive aspects. By contrast, when Bornmann, Wagner, and Leydesdorff (2018) looked at country shares, they focused on the shares of domestic publications that are cited rather than shares of global citations acquired. They analyzed which national publications were cited in the global slice of "elite publications" (defined as the 1% most frequently-cited publications). This is a "backward citing" view while conventional analysis of citation shares – impact – is based on a "forward citing" view; in the backward view, the analysis is directed towards the shoulders on which (impactful) research subsequently stands (Bornmann, de Moya-Anegón, & Leydesdorff, 2010; Merton, 1965). The results of Bornmann et al. (2018) confirm a continuing significance for US research, in agreement with Rodríguez-Navarro and Narin (2018). Further "strong shoulders" are Switzerland, the Netherlands, the UK, and Sweden. Although Germany is often identified as strong in terms of citation impact (e.g. Bornmann & Leydesdorff, 2013; Marshall & Travis, 2011), the results of Bornmann et al. (2018) suggest that Germany does not belong to the group of top-performing countries.

In the present paper we have two objectives, addressing both data content and statistical research in the context of the previously published literature. Following the approach of Bornmann et al. (2018), we investigate the shoulders on which national research stands. We compare Germany with two other leading European nations. In a recent study, Bakare and Lewison (2017) show that "there is a clear tendency for authors of scientific papers to over-cite the papers by their fellow countrymen (and countrywomen) relative to the percentage presence of their papers in world output in the same field" (p. 1199). These authors investigated six scientific fields (astronomy, birds, cancer, chemistry, diabetes and engineering) and seven publication years (1980, 1985, 1990, 1995, 2000, 2005 and 2010). The tendency for "over-citation" is stronger for scientific fields of more national interest and has decreased over time. Tang, Shapira, and Youtie (2014) choose the term of "clubbing effect" as a label for a similar phenomenon.

Based on a similar cited references analyses to those used by Bornmann et al. (2018), we investigate in this study whether these differentials in citing patterns are found when we focus on three leading scientific nations.

The second objective is to investigate the relationship between citation performance and country assignments of papers. We used regression models to target this objective by controlling for moderating variables (e.g. the number of countries to which the co-authors of a cited paper belong). The regression analysis follows an approach introduced by Bornmann (2016). National data used in this study were not analysed on an aggregated country-level (as it is usually the case in the studies discussed above) but on the level of single cited references. The chief advantage of such a methodology is that one can show the change in importance of a cited reference in "elite publications" relative to all other cited references in a given year of citing papers.

# 3 Methods

## 3.1 Dataset

This study is about the shoulders on which highly-cited research stands. The precursor study by Bornmann et al. (2018) is based on all cited references in the global set of papers that belong to the 1% most frequently cited papers ("highly-cited papers"). Citation counts increase over time at a field-dependent rate, so raw counts are normalized for time and subject category. Percentiles of citation counts are used as a proxy for the quality of the cited publication (Bornmann & Haunschild, 2017). The calculation of percentiles is an established method in bibliometrics (Hicks, Wouters, Waltman, de Rijcke, & Rafols, 2015). Among the various ways to compute citation percentiles, we use in this study Hazen percentiles (Hazen, 1914):[1] higher percentiles indicate higher citation impact (Bornmann, Leydesdorff, & Mutz, 2013). For example, a percentile value of 99 indicates that the paper belongs to the 1% most frequently cited papers in its publication year and subject category.

The bibliometric data used in this paper are from an in-house database developed and maintained by the Max Planck Digital Library (MPDL, Munich) which is based on the Web of Science (WoS, Clarivate Analytics). We included only papers from the natural sciences in our study (using the OECD field categorization scheme on the major code level) because of the following reasons: (1) we expect differences in the results between (i) natural sciences and (ii) social sciences and humanities. (2) In many social sciences and humanities fields, it is problematic to use bibliometrics. The analytical set of papers was reduced by focusing on all cited references in the national publication set for three countries: Germany plus the Netherlands, and the UK as comparable and high-performing countries. To evaluate change over time, we include papers published in the sample countries from three publication years:

---

[1] Hazen percentiles are calculated using the formula $(100 * (i-0.5)/n)$ where $n$ is the total number of papers in a specific field and publication year, and $i$ the rank number of papers (when the papers are sorted in descending order by citation impact). Papers with equal citation counts are assigned the average rank.

2004, 2009, and 2014. The choice of 2014 as the most recent publication year was dictated by the need to maintain a citation window of at least three years for publication analyses (Glänzel & Schöpflin, 1995). The current release of available data includes citations through to the end of 2016.

Citation rates vary for publications of different document type (Aksnes, 2006) and subject categories. For this reason, only documents classified as "journal articles" were included in the study. Articles from the three publication years were classified bimodally as belonging either to (1) the 1% most-frequently cited, by subject category and publication year, or (0) not being so highly-cited (thus producing a binary variable where 0 = not highly-cited and 1 = highly-cited).

The shoulders on which published articles stand are the references they cite. In this study, only cited references could be considered which are themselves covered in the WoS. Metadata, such as the countries of the authors, are only available for these cited references. The cited references for this study are also restricted to articles to avoid distortions by introduced by including different document types, a restriction that has the additional effect of slightly reducing the number of cited references. To standardize comparisons across the three sample years, cited references are analyzed only from the most recent three years for each instance: publication year 2004 takes cited reference years for 2001-2003; publication year 2009 takes cited references for 2006-2008; publication year 2014 takes cited references for 2011-2013. Thus, the shoulders are defined by the most recent articles prior to publication. This accords with the tendency of citing authors to include publications from the recent years in their reference lists (Bornmann, Ye, & Ye, 2017). In other words, we focus on the research front and fields with a relatively short-term citation cycle (de Solla Price, 1970; Leydesdorff, Bornmann, Comins, & Milojević, 2016).

The final counts of data points included in the statistical analyses are shown in the relevant results sections.

**3.2 Statistical model**

We used regression models to analyze cited references as basic units. The dependent and independent variables refer to the cited paper (e.g. the authors' country) and the citing paper (e.g. whether it is highly-cited or not). Logistic rather than linear regression is used because the dependent variable is a binary variable: the units $i$ are references cited in articles that are themselves highly-cited ($y_i = 1$) or not ($y_i = 0$). Since we are interested in the effect of the characteristics of cited references (especially their country assignments) on the probability of being highly-cited, the dependent variable is a binary variable: it is the status of the citing article being highly-cited or not. The link between the observed binary variable $y_i$ and the continuous latent variable $y_i^*$ in the regression model is defined as

$$y_i = \begin{cases} 1 \text{ if } y_i^* > 0 \\ 0 \text{ if } y_i^* \leq 0 \end{cases} \quad (1)$$

References cited in highly-cited articles ($y_i = 1$) are cases with positive values of $y^*$; references cited in articles that are not highly-cited ($y_i = 0$) are cases with negative or null values of $y^*$ (Long & Freese, 2014).

In the multiple logistic regression the dependent variable (here: the probability of citing articles being highly-cited) is predicted by a linear combination of several independent variables (here: especially the country assignments of cited references) (Kohler & Kreuter, 2012):

$$y_i^* = \beta_0 + \beta_1 x_{1i} + \beta_2 x_{2i} + \cdots + \beta_{k-1} x_{k-1,i} + \varepsilon_i \quad (2)$$

In equation 2, $x_{1i}$ is the value of the first independent variable for cited reference $i$, $x_{2i}$ is the corresponding value for the second independent variable etc. $\beta_1, \beta_2, \ldots, \beta_{k-1}$ are the regression parameters which represent the weights assigned to the independent variables. $\beta_1, \beta_2, \ldots, \beta_{k-1}$ are unknown constants which are estimated from the underlying data; $\varepsilon_i$ is chance variation (i.e. noise, disturbance, or error) (Hoaglin, 2016). To simplify the interpretation of the results for the independent variables, the percentages of change in odds are given which are calculated by using the formula:

$$100\{\exp(\delta\beta_k) - 1\} \tag{3}$$

This percentage expresses the practical significance of the results (Cumming, 2012).

We included independent variables on the level of citing as well as cited papers: (1) On the level of the citing paper, we include the publication year to differentiate between the cohorts of citing years. Furthermore, we consider the number of author-countries listed on a paper as independent variable. In this study, we define papers with at least one German author (an author with a German affiliation) as a German paper and define similarly for the UK and the Netherlands. Since many papers in the natural sciences are produced through international co-authorships, it is necessary to control the number of mentioned countries on the citing paper in the regression models (see here Abramo & D'Angelo, 2018). Iribarren-Maestro, Lascurain-Sanchez, and Sanz-Casado (2007) suggest the number of countries co-authoring a paper is positively related to its citation impact.

(2) On the level of cited references, we consider the number of countries mentioned on a cited paper. The reason is the same as for the consideration of this variable on the citing side. The year of publication considers the time difference between publication of the citing paper and the cited paper in the regression model (e.g. a value of "2" for the independent variable means that the cited reference was published two years before the citing publication). The citation percentile indicates the citation impact of the cited paper. We control the quality of the cited paper in a part of the regression models to see whether country effects are visible independent of the quality of the cited paper. The language of the citing paper has been included as independent variable, since one could expect a stronger national orientation in citing for national than for international papers (in the case of German and Dutch citing papers). Furthermore, ten countries of particular research significance are, when mentioned among the addresses of a cited paper (or not), considered as binary independent variables in the models. Since the country variables in the various models (which we performed) differ, we explain the variables in more detail in section 4.

We performed robustness checks to analyze whether the results of the regression models significantly changed dependent on the included countries as independent variables. First, we re-calculated all regression models including the five most important (most frequently cited) countries (instead of the ten) to investigate how this affects our results. Second, we added randomly five and ten further countries, respectively, from the countries at the eleventh and fortieth place among the most frequently cited ones, respectively. The results of these further analyses are added to the appendix. The robustness checks for all models showed that the value of the coefficients may change (slightly), but the sign of the coefficients did not change. There is only one exception: the sign of the coefficient for UK changed from positive to negative in one model (see Table 12).

It is a further requirement of regression analysis that the cases in the dataset are independent from one another. In the dataset of this study, this requirement is violated by including more than one cited reference from an individual citing article (whether or not they were highly-cited). In other words, the cited references are clustered within citing articles. Stata, the program used for the statistical analyses (StataCorp., 2017), provides the option of computing robust standard errors with additional corrections for the effects of clustered data (Long & Freese, 2014). The use of this option in the regression model does not change the coefficient estimates, but the standard errors. A detailed discussion of logistic regressions with clustered data can be found in Hosmer, Lemeshow, and Sturdivant (2013). According to Angeles, Cronin, Guilkey, Lance, and Sullivan (2014), Stata's cluster option "is a post-estimation modification (meaning that it influences standard error estimation, but only after point estimation is complete). The standard error estimates generally grow larger because the correlation of errors at the individual level left the misimpression in the first regression (which ignored possible correlation of errors) of more independent variation in Y between observations than was actually the case" (p. 11).

# 4  Results

Regression models calculated for three countries (Germany, the Netherlands, and the UK) are used to explore the effect of cited references from specific countries on the probability of being highly-cited. The results are presented in sections 4.1, 4.2, and 4.3.

**4.1  Germany**

The underlying data for the regression analyses for Germany are articles (whether or not they were highly-cited) with at least one German affiliation. The cited references in these articles have been published by authors from well over 100 countries, but the regression analysis would be unwieldy with so many countries so the sub-set included as independent variables are only those that are relatively frequently referenced. Table 1 shows the ten countries that have the highest reference frequency. Where a cited reference is assigned to more than one country (i.e. it has author addresses for more than one country), then the contributing countries have been fractionally counted. For example, if the authors of the cited paper are from three countries, the paper has been assigned one-third to each country (regardless of the frequency of authors per country for that paper).

Table 1. Countries most frequently referenced in **German** articles published in 2004, 2009, and 2014. The number for each country is an average value across the number of cited references in articles from the three citing years. Only the cited references from the last three years are considered (see section 3.1).

| Country | Fractional counting |
|---|---|
| United States | 80,542.01 |
| Germany | 78,674.35 |
| UK | 20,587.75 |
| France | 15,314.00 |
| Japan | 14,283.49 |
| China | 12,544.99 |
| Italy | 9,159.25 |
| Canada | 8,457.97 |

| Switzerland | 7,982.81 |
| The Netherlands | 7,482.90 |

The regression analysis for Germany is based on 973,296 cited references (cited articles) in 129,958 citing articles. The number of cited references is slightly reduced (by 189 items) by including only those cases with no missing values for all variables. There is an average of 7.5 cited references (minimum=1, maximum=196). Table 7 (in appendix 6.1) presents statistics for the dependent and independent variables included in the regression analysis. In the case of binary variables, the mean can be interpreted as proportion. For example, the dependent variable is a binary variable: 3% of the references are cited in German papers that are among the 1% most frequently cited papers. The variables in Table 7 refer to two levels: citing and cited (see the description of the independent variable in section 3.2).

Table 2 shows the results of two regression analyses (models A and B). The only difference between the models is the inclusion of the citation percentiles as an independent variable in model B. The effect on the citation impact of the citing paper of author country-count on the cited papers has been tested by excluding (model A) and including (model B) the cited paper's impact: do the results change if citation impact as a proxy of cited paper quality is included? Table 2 lists the coefficients in log-odds units, which are the values for the regression equation (see section 3.2) for predicting the dependent variable. This column also contains the 95% confidence intervals (CIs) for the odds ratios. Since the sample size of this study is very high, statistical significance of the coefficients is scarcely meaningful (Kline, 2004).

The relationship between the independent and dependent variables can be interpreted based on the percentage changes in Table 2. The percentage for the number of countries, for example, can be interpreted as follows: for each additional country among the addresses of a cited paper, the odds for the citing paper of being highly-cited decrease by 7.8% where all other variables are held constant. This percentage slightly changes if paper percentiles are included in the model (from 7.8% to 7.2%). Both results mean that the likelihood that the citing paper will be highly-cited is affected by the number of countries attributed to the cited reference. Since the percentages are of the same order in models A and B, there is also little effect from the citation impact of the cited references.

Table 2. Odds ratios with 95% confidence intervals (CIs) and percentage changes in odds for unit increase in the independent variables as the results of two regression analyses (excluding and including the cited paper's impact percentile as independent variable).

|  | Model A: Excluding paper percentile | | Model B: Including paper percentile | | Factor change in percentages |
|---|---|---|---|---|---|
| Variable | Odds ratio, 95% CI | Percentage change | Odds ratio, 95% CI | Percentage change | |
| Citing paper level | | | | | |
| Publication year | | | | | |
| 2009 | 1.15 | 15.5 | 1.14 | 14.5 | 1.07 |
|  | [0.98,1.36] | | [0.97,1.35] | | |
| 2014 | 1.47*** | 47.2 | 1.43*** | 42.5 | 1.11 |
|  | [1.25,1.73] | | [1.21,1.67] | | |
| Number of countries | 1.12*** | 11.8 | 1.12*** | 11.6 | 1.02 |
|  | [1.10,1.14] | | [1.10,1.14] | | |
| Cited reference level | | | | | |
| County | | | | | |
| USA | 1.52*** | 51.9 | 1.24*** | 24.1 | 2.15 |
|  | [1.42,1.63] | | [1.16,1.33] | | |
| Germany | 0.89*** | -11.1 | 0.89*** | -11.4 | 0.97 |
|  | [0.84,0.94] | | [0.83,0.94] | | |
| UK | 1.37*** | 37.3 | 1.23*** | 23.3 | 1.60 |
|  | [1.28,1.47] | | [1.15,1.32] | | |
| France | 1.16*** | 16.4 | 1.11* | 10.6 | 1.55 |
|  | [1.08,1.26] | | [1.02,1.19] | | |
| Japan | 1.10* | 10.1 | 1.05 | 5.2 | 1.94 |
|  | [1.01,1.20] | | [0.97,1.14] | | |
| China | 1.23** | 22.7 | 1.17* | 16.8 | 1.35 |
|  | [1.07,1.40] | | [1.02,1.33] | | |
| Italy | 1.17*** | 16.9 | 1.15** | 15.3 | 1.10 |
|  | [1.07,1.28] | | [1.05,1.26] | | |
| Canada | 1.33*** | 32.7 | 1.26*** | 26 | 1.26 |

| Variable | Model A: Excluding paper percentile | | Model B: Including paper percentile | | Factor change in percentages |
|---|---|---|---|---|---|
| | Odds ratio, 95% CI | Percentage change | Odds ratio, 95% CI | Percentage change | |
| | [1.21,1.46] | | [1.15,1.38] | | |
| Switzerland | 1.03 | 2.5 | 0.91 | -9.3 | |
| | [0.92,1.14] | | [0.82,1.01] | | |
| The Netherlands | 1.20*** | 20.3 | 1.10 | 10.2 | 1.99 |
| | [1.09,1.33] | | [1.00,1.22] | | |
| Years back of cited reference year | | | | | |
| 2 | 0.91*** | -9 | 0.91*** | -9 | 1.00 |
| | [0.87,0.95] | | [0.87,0.95] | | |
| 3 | 0.78*** | -21.7 | 0.76*** | -23.8 | 0.91 |
| | [0.75,0.82] | | [0.73,0.80] | | |
| Number of countries | 0.92*** | -7.8 | 0.93*** | -7.2 | 1.08 |
| | [0.90,0.94] | | [0.91,0.95] | | |
| English paper | 2.17* | 117.2 | 0.86 | -14.3 | |
| | [1.09,4.34] | | [0.44,1.68] | | |
| Citation percentile | | | 1.04*** | 3.8 | |
| | | | [1.03,1.04] | | |
| Observations | 973,296 | | 973,296 | | |

Notes.
Exponentiated coefficients; 95% confidence intervals in brackets.
* $p < 0.05$, ** $p < 0.01$, *** $p < 0.001$

If we consider the specific countries in Table 2, there are high percentage values for the USA (51.9%), the UK (37.3%), and Canada (32.7%) (again, irrespective of the paper percentiles of the cited references). The implication is that where published research from these countries is cited as their "shoulders", then it is more likely that the citing paper will be highly-cited. By contrast, with 2.5%, Switzerland's output has a significantly lesser effect on being highly-cited. It is notable that the likelihoods for most countries are significantly reduced (see the column factor change in percentages) if impact percentiles are included in the model. By including this variable, the country effect becomes visible independently of the quality of the papers. Model B shows the effect where cited paper impact is controlled. The changes for the USA (reduced by a factor of around 2) and for the UK and France (reduced by a factor of around 1.6) reflect the research capacity of these countries. The output of these countries may be of high quality (citation impact), but the residual country variables in model B demonstrate the varying likelihood that these outputs will contribute to highly-cited papers. Reasons may include differences in the availability of papers for citation, self-citation effects, and – most importantly –relative reputation (of countries or institutions) in their global network.

Germany is the only country with a negative percentage change in both models in Table 2. This means the likelihood that the citing paper is highly-cited decreases by about 11% if that paper has cited at least one German paper. Figure 1 shows the different probabilities that the article will be highly-cited related to German authorship on the cited publication and the publication year of the citing article. The trend suggests the effect, with or without German authors, increases over the citing years. The further calculation of ideal types (which are hypothetical observations with substantively illustrative values, see Long & Freese, 2014) shows that the estimated probability of a German citing paper being highly-cited is 0.027 where a German paper is cited (the results are not shown in a table). This accords with the mean value of being highly-cited in the dataset (see Table 7) and is significantly lower than, for example, the estimated probability for the USA (0.036). Taken together, the results for Germany show that its presence among cited papers is negatively related to the likelihood of a German citing paper being one that is highly-cited. Furthermore, the citation of German papers does not seem to be triggered by quality aspects: the percentage change between models A and B in Table 2 is relatively small.

Switzerland is the only country in Table 2 with a positive (model A) and negative (model B) percent change: if quality aspects are excluded from the regression model the coefficient is small, but positive (2.5); if Hazen percentiles are included in the model, the percent change becomes negative (-9.3). There seems to be a bias among authors of German papers to cite Swiss papers independent of the quality of the cited papers which leads to the lower probability of being highly-cited. One reason for this result might be that many researchers from Germany have worked in the German-speaking part of Switzerland and maintain a close collaboration.

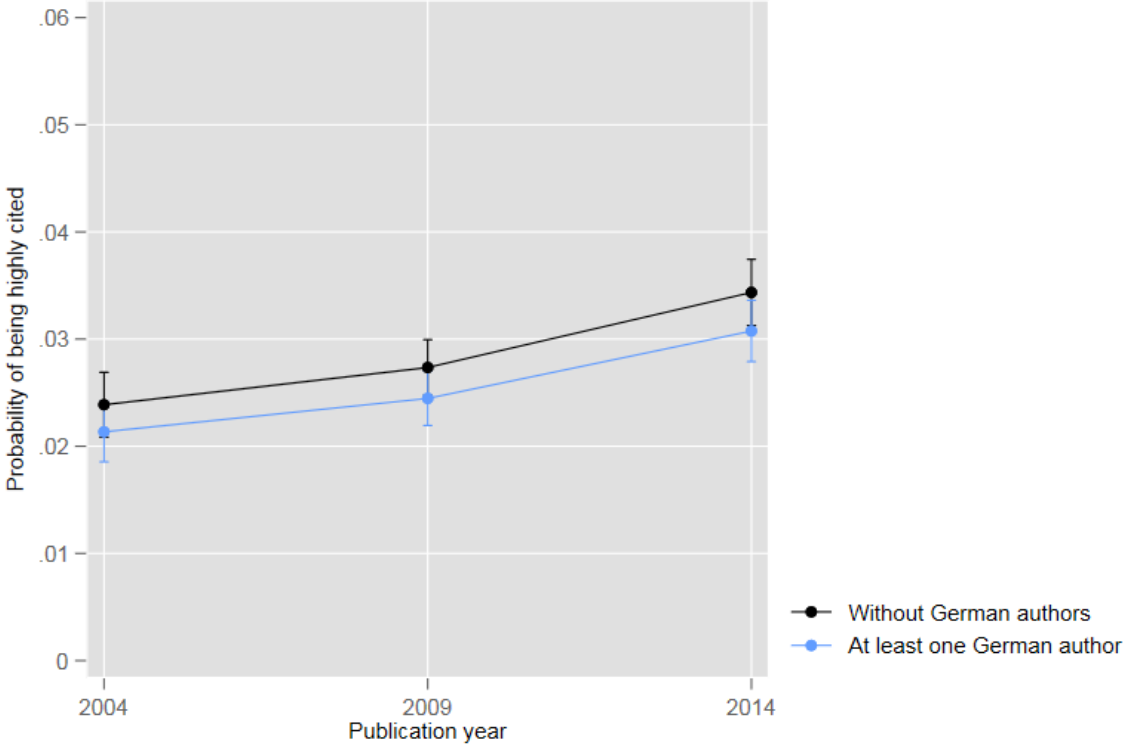

Figure 1. Likelihood of being highly-cited (from regression model A) related to the publication year of the citing paper and the presence of a German-authored article among its cited references.

Our result concerning Germany appear to agree with Bornmann et al. (2018): this country does not belong to the group of top-nations if we address the shoulders on which research stands. However, the results for Germany are different from the results for nearly all other countries (with the exception of Switzerland) in both models not in scale but in sign: that is to say, the outcome for the set of citing papers with a German address suggests that cited articles with a German address are instrumentally linked to the result of the citation impact analysis. In the following sections, the comparison with the Netherlands (section 4.2) and the UK (section 4.3) will show whether this is a common phenomenon across national systems.

**4.2   The Netherlands**

The countries most frequently referenced articles in articles with at least one Netherlands-author and published in 2004, 2009, and 2014 (see Table 3) are similar to those for Germany (see Table 1). The only change is that Switzerland has been replaced by Spain.

Table 3. Countries which have been most frequently referenced in the **Dutch** papers published in 2004, 2009, and 2014. The number for each country is an average value across the number of cited references in papers from the three years. Only the cited references from the last three years are considered (see section 3.1).

| Country | Fractional counting |
|---|---|
| United States | 23,468.12 |
| The Netherlands | 14,206.90 |
| Germany | 7,468.53 |
| UK | 7,030.92 |
| France | 4,269.06 |
| Japan | 3,114.56 |
| China | 2,752.61 |
| Italy | 2,632.44 |
| Canada | 2,574.41 |
| Spain | 2,272.94 |

The regression analyses for the Netherlands included 266,000 cited references in 34,368 citing articles, with an average around 8 cited references (minimum=1, maximum=315). The only cases considered in the analyses were those with no missing values for any variables (60 cases were excluded). Table 8 (in appendix 6.2) shows the key figures for the dependent and independent variables included in the regression models. The results show that 26% of the articles cited by Netherlands authors can be assigned to the Netherlands, which is 15 points lower than the articles for the USA (with 0.41 or 41%) which again heads the table.

The results of the regression analyses are presented in Table 4: model A excludes and model B includes percentiles of the cited paper. The results for Germany no longer shows an odds ratio smaller than 1: when seen from the perspective of Netherlands authorship, German performance is not different from that of other countries.

Table 4. Odds ratios with 95% confidence intervals (CIs) and percentage changes in odds for unit increase in the independent variables as the results of two regression analyses (excluding and including the paper percentile as independent variable)

| Model A: Excluding paper percentile | | | Model B: Including paper percentile | | |
|---|---|---|---|---|---|
| Variable | Odds ratio, 95% CI | Percentage change | Odds ratio, 95% CI | Percentage change | Factor change in percentages |
| Citing paper level | | | | | |
| Publication year | | | | | |
| 2009 | 1.41** | 41.4 | 1.40** | 39.8 | 1.04 |
|  | [1.11,1.80] |  | [1.10,1.78] |  |  |
| 2014 | 1.82*** | 81.8 | 1.78*** | 77.7 | 1.05 |
|  | [1.42,2.33] |  | [1.38,2.29] |  |  |
| Number of countries | 1.10*** | 10.4 | 1.10*** | 10.3 | 1.01 |
|  | [1.08,1.13] |  | [1.08,1.13] |  |  |
| Cited reference level | | | | | |
| Country | | | | | |
| USA | 1.46*** | 45.6 | 1.26*** | 25.8 | 1.77 |
|  | [1.31,1.62] |  | [1.14,1.39] |  |  |
| The Netherlands | 0.85** | -14.8 | 0.86** | -14.0 | 1.06 |
|  | [0.77,0.94] |  | [0.78,0.95] |  |  |
| Germany | 1.06 | 6.4 | 1.01 | 0.6 | 10.67 |
|  | [0.95,1.19] |  | [0.90,1.12] |  |  |
| UK | 1.29*** | 28.6 | 1.20*** | 19.6 | 1.46 |
|  | [1.16,1.42] |  | [1.08,1.32] |  |  |
| France | 1.08 | 7.7 | 1.05 | 5.3 | 1.45 |
|  | [0.96,1.21] |  | [0.94,1.19] |  |  |
| Japan | 1.03 | 2.5 | 1.01 | 0.8 | 3.13 |
|  | [0.91,1.16] |  | [0.89,1.14] |  |  |
| China | 1.16 | 15.6 | 1.16 | 15.9 | 0.98 |
|  | [0.87,1.53] |  | [0.87,1.54] |  |  |
| Italy | 1.19** | 18.7 | 1.20** | 19.6 | 0.95 |
|  | [1.04,1.35] |  | [1.05,1.36] |  |  |
| Canada | 1.31*** | 31.0 | 1.27** | 27.0 | 1.15 |
|  | [1.14,1.51] |  | [1.10,1.47] |  |  |
| Spain | 1.09 | 9.0 | 1.12 | 12.1 | 0.74 |
|  | [0.95,1.25] |  | [0.97,1.29] |  |  |
| Years back of cited reference year |  |  |  |  |  |
| 2 | 0.92** | -8.3 | 0.92** | -7.8 | 1.06 |
|  | [0.86,0.98] |  | [0.87,0.98] |  |  |
| 3 | 0.81*** | -19.3 | 0.79*** | -20.6 | 0.94 |
|  | [0.75,0.87] |  | [0.74,0.86] |  |  |
| Number of countries | 0.92*** | -7.7 | 0.92*** | -7.9 | 0.97 |
|  | [0.90,0.95] |  | [0.90,0.95] |  |  |
| English paper | 4.25 | 325.4 | 2.02 | 101.6 | 3.20 |
|  | [0.82,22.16] |  | [0.38,10.78] |  |  |
| Citation percentile |  |  | 1.03*** | 2.7 |  |
|  |  |  | [1.02,1.03] |  |  |
| Observations | 266,000 |  | 266,000 |  |  |

Notes.
Exponentiated coefficients; 95% confidence intervals in brackets.
* $p < 0.05$, ** $p < 0.01$, *** $p < 0.001$

The Netherlands has an odds ratio smaller than 1, unlike the results for Germany. In other words, the change from the German citing set to the Netherlands citing set leads to a change between Germany and the Netherlands in the odds ratio associated with the same-country cited set. As for Germany (see Table 2), inclusion of percentiles as a proxy for quality has scarcely an effect on the outcomes of the regression analyses (the percentage change values for the Netherlands are almost equal in both models – -14.8 and -14.0, see Table 4).

Figure 2 shows how the likelihood that a citing paper will be highly-cited (from regression model A) varies with the presence or absence of Dutch authorship on the cited publication and the publication year of the citing paper. The results look similar to the German results (see Figure 1). For all citing years, the probability of being highly-cited is lower where cited papers have at least one Dutch author and the difference between groups increases in more recent years.

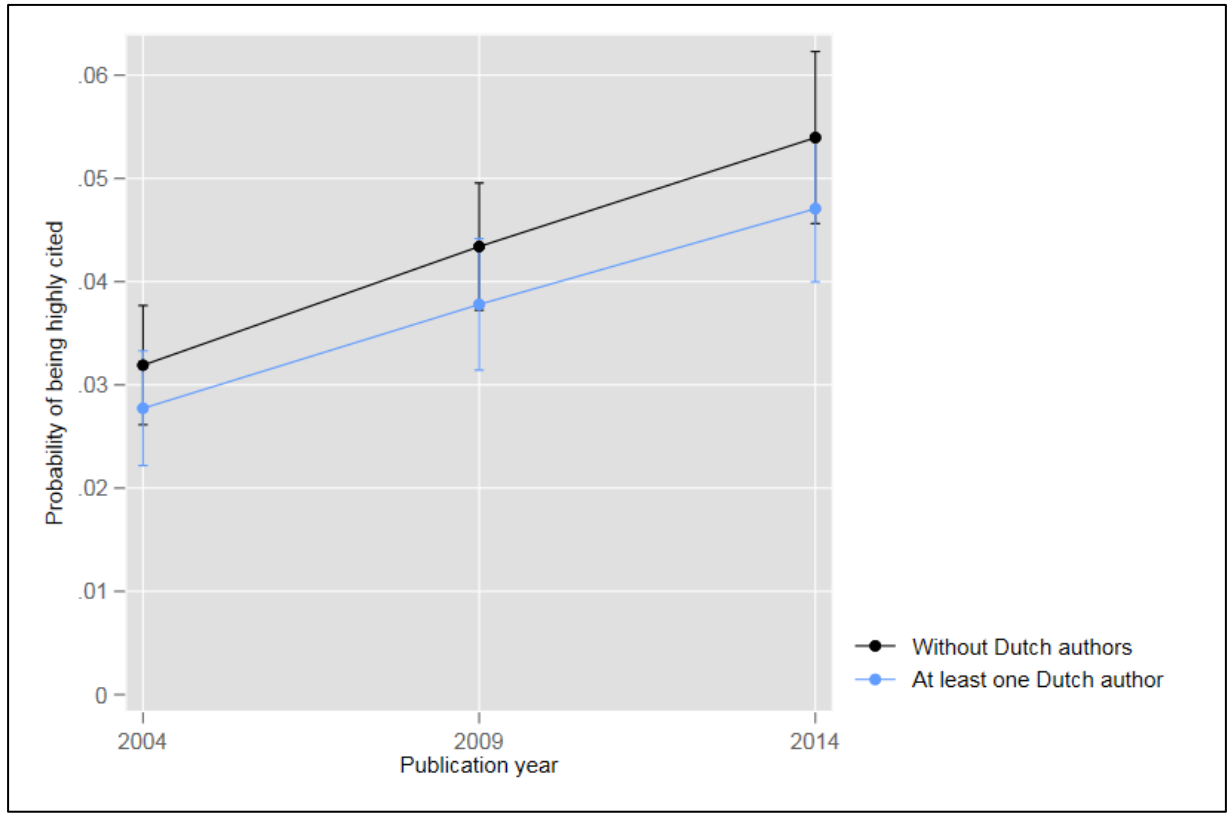

Figure 2. The likelihood of a Netherlands-authored paper being highly-cited (from regression model A) with regard to the presence or absence of a Dutch author on a cited publication and the publication year of the citing paper.

### 4.3 UK

The UK analysis reiterates the approach applied to Germany and the Netherlands. The countries included in the UK regression models (Table 5) are similar to those for the Netherlands (see Table 3). The UK statistical analysis is based on 840,697 cited articles in 111,454 citing articles with at least one author with a UK address. There were 183 cases excluded from the analyses because of missing values; each UK citing article has an average of about 7 cited references (minimum=1, maximum=315).

Table 5. Author countries most frequently referenced in citing articles with a UK author and published in 2004, 2009, and 2014. The number for each country is an average value across the number of cited references in articles from the three years. Only the cited references from the last three years are considered (see section 3.1).

| Country | Fractional counting |
|---|---|
| United States | 76,207.62 |
| UK | 60,145.82 |
| Germany | 19,788.69 |
| France | 13,019.26 |
| Japan | 10,714.03 |
| China | 10,486.93 |
| Canada | 8,823.48 |
| Italy | 7,910.31 |
| Spain | 6,946.06 |
| Australia | 6,872.21 |

Table 9 (in appendix 6.2) presents the key values for the dependent and independent variables included in two regression analyses. The mean citation percentile (last row) displays about the same values as for Germany (see Table 7) and the Netherlands (see Table 8). In other words, the authors from all three countries cite papers from the last three years that have a similar level of citation impact.

The results of two regression models A and B do not repeat the pattern observed for Germany and the Netherlands (see Table 6). The UK has an odds ratio, which is close to one, but not negative. The percentage values of the outcomes of both regression analyses are lower for the UK than for many other countries in the table.

Table 6. Odds ratios with 95% confidence intervals (CIs) and percentage changes in odds for unit increase in the independent variables as the results of two regression analyses (excluding and including the paper percentile as independent variable)

| Model A: Excluding paper percentile | | | Model B: Including paper percentile | | |
|---|---|---|---|---|---|
| Variable | Odds ratio, | Percentage | Odds ratio, | Percentage | Factor |

|  | 95% CI | change | 95% CI | change | change in percentages |
|---|---|---|---|---|---|
| Citing paper level | | | | | |
| Publication year | | | | | |
| 2009 | 1.22* | 21.5 | 1.20* | 19.9 | 1.08 |
|  | [1.05,1.41] |  | [1.03,1.39] |  |  |
| 2014 | 1.52*** | 51.7 | 1.46*** | 46.0 | 1.12 |
|  | [1.31,1.76] |  | [1.26,1.69] |  |  |
| Number of countries | 1.11*** | 10.9 | 1.11*** | 10.7 | 1.02 |
|  | [1.09,1.13] |  | [1.09,1.13] |  |  |
| Cited reference level | | | | | |
| Country | | | | | |
| USA | 1.58*** | 58.0 | 1.31*** | 31.1 | 1.86 |
|  | [1.48,1.68] |  | [1.23,1.40] |  |  |
| UK | 1.04 | 4.3 | 1.02 | 2.0 | 2.15 |
|  | [0.99,1.10] |  | [0.97,1.08] |  |  |
| Germany | 1.15*** | 14.9 | 1.05 | 5.1 | 2.92 |
|  | [1.07,1.23] |  | [0.98,1.13] |  |  |
| France | 1.09* | 9.3 | 1.04 | 4.3 | 2.16 |
|  | [1.01,1.18] |  | [0.97,1.12] |  |  |
| Japan | 1.16*** | 16.0 | 1.14** | 13.6 | 1.18 |
|  | [1.07,1.26] |  | [1.05,1.23] |  |  |
| China | 1.31*** | 30.7 | 1.26*** | 26.4 | 1.16 |
|  | [1.14,1.50] |  | [1.10,1.45] |  |  |
| Canada | 1.30*** | 29.7 | 1.23*** | 23.0 | 1.29 |
|  | [1.20,1.41] |  | [1.13,1.33] |  |  |
| Italy | 1.02 | 2.2 | 1.01 | 1.3 | 1.69 |
|  | [0.94,1.11] |  | [0.93,1.10] |  |  |
| Spain | 1.16** | 16.2 | 1.17** | 17.3 | 0.94 |
|  | [1.05,1.28] |  | [1.06,1.29] |  |  |
| Australia | 1.42*** | 41.8 | 1.33*** | 32.5 | 1.29 |
|  | [1.29,1.56] |  | [1.20,1.46] |  |  |
| Years back of cited reference year | | | | | |
| 2 | 0.92*** | -8.1 | 0.92*** | -8.3 | 0.98 |
|  | [0.89,0.95] |  | [0.88,0.95] |  |  |
| 3 | 0.78*** | -22.0 | 0.76*** | -23.9 | 0.92 |
|  | [0.74,0.82] |  | [0.73,0.80] |  |  |
| Number of countries | 0.92*** | -8.1 | 0.92*** | -8.0 | 1.01 |
|  | [0.90,0.94] |  | [0.90,0.94] |  |  |
| English paper | 1.41 | 41.2 | 0.61 | -39.2 |  |
|  | [0.59,3.39] |  | [0.26,1.42] |  |  |
| Citation percentile |  |  | 1.04*** | 3.5 |  |
|  |  |  | [1.03,1.04] |  |  |

| Observations | 840,697 | | 840,697 | | |

Notes.
Exponentiated coefficients; 95% confidence intervals in brackets.
$^{*} p < 0.05$, $^{**} p < 0.01$, $^{***} p < 0.001$

This difference for UK-citing articles compared to those of the other countries (Germany and the Netherlands) can be seen in Figure 3. The probability of being highly-cited is even slightly higher for papers referencing papers with at least one British author than for papers referencing papers without British authors.

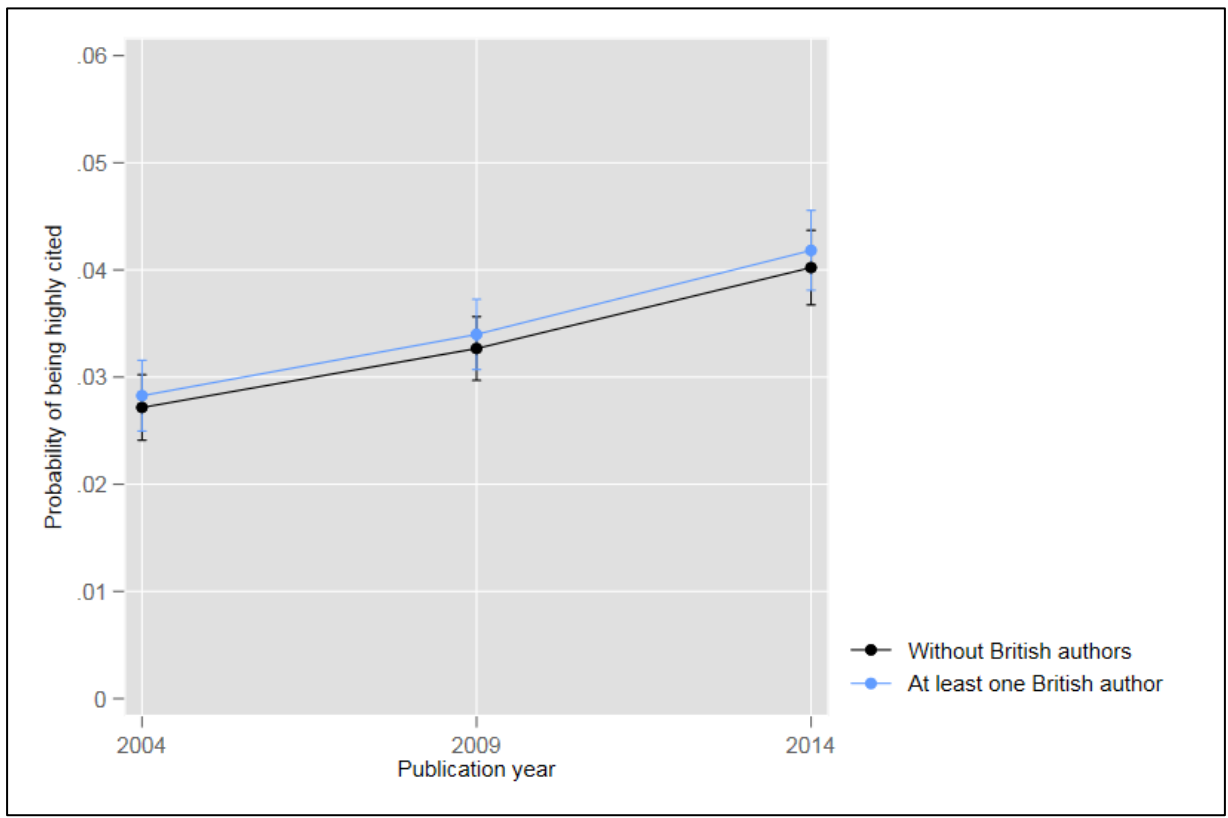

Figure 3. Predictions of being highly-cited (from regression model A) in dependence of the presence of a British author on the cited publication and the publication year of the citing paper

# 5  Discussion

Internationally co-authored research papers are becoming more frequent and they account for the more highly-cited component of the output of leading research economies. The authorship of that output is, by definition, shared with other countries; much less is known about the research on which this highly-cited international network draws for its inspiration and authority.

In this paper we have shown for two countries (Germany and the Netherlands), that articles co-authored by researchers in one of these countries are less likely to be among the globally most highly-cited if they also cite "domestic" research (i.e. research authored by the same country). To put this another way, papers which stand on domestic shoulders are expected to be less well-cited than papers that stand on international shoulders. The results also point to a national bias in citing (see section 2) which has been explained by Gingras and Khelfaoui (2018) as follows: national publications are "more visible in their country than international publications, in all disciplines" (p. 525). However, the difference in citing between "domestic" and "international" research is not visible for the UK highly-cited papers.

As with most analyses of research activity and performance, the comparative aspects are of particular policy significance. The US is an important "shoulder" for other country's domestic and international literature. Similar findings have been published by Gingras and Khelfaoui (2018). The differential analysis between more and less highly-cited papers reveals a significantly greater likelihood that research citing domestic literature will be less well cited for Germany and the Netherlands. These effects are consistent over the three sample years. By implication, this may suggest that there is a potential separation between the domestic and internationally-engaged parts of the research base in those countries. Policy attention in the countries may need to focus on the location, the causes, and on measures that mitigate further disengagement.

The evidence for Germany and the Netherlands suggests that citing papers that draw on the international literature are more likely to attract citations. This could be for two reasons: first, they cite international literature because they contain analysis and results of wide significance, which in turn attracts attention; or, second, because they also have diverse international authorship and come to the direct attention of a wider readership. Equally significant is the finding of our study that less well-cited research draws more heavily on domestically-authored references.

Our results suggest a stratification in national publication systems in terms of international or domestic orientation of the knowledge base from which authors cite. This stratification is not (or no longer) an effect of language because international publications in Dutch are virtually non-existent. Funding is a source of national orientation which may distract from international participation as may be other elements of the institutional structures of the publication system which are culturally and nationally specific. The role of the German-speaking part of Switzerland which functions as an independent part of the German system is worth further investigation. Significantly, the UK is fully integrated in the Anglophone publication system.

A domestic orientation is not necessarily an indicator of weak achievement. Indeed, it may be an important component in the research process that the domestic development of the research "platform" on which "peak" activity can build should be more domestically referential. This point is addressed by Rochmyaningsih (2017) in a similar form, but in another context: "In the case of Indonesia, the small number of papers contains research about climate change, earthquake geology, the genetics of malaria, tropical forests, peatlands and high-energy physics. The findings could help to make our country a better place to live in. Yet, in my experience in covering science issues, most of these findings are ignored in the process of policymaking. This is a problem that Indonesia – and others – should address. For example, last month, Indonesian scientists published a study of a new geological fault system in the Indian Ocean, which increases the chance of earthquakes in the north of Sumatra (S. C. Singh *et al. Sci. Adv.* **3,** e1601689; 2017). The value of such a paper is not in its contribution to boosting our national scientific profile abroad, but in its role in improving disaster-mitigation policy at home" (p. 7). We think that this point requires further exploration.

Why is this important for research policy and management? If internationally collaborative research is increasingly the "location" of the most significant work then research that is not founded on this (i.e. that does not cite this) may be less well-informed. It is not standing on the "shoulders of giants", it does not see further and it is therefore less likely to be of wider significance in the international literature. The UK articles' probability of being highly-cited, where there is a very small difference between articles that do and do not stand on domestic shoulders, is competitive with that of the US. To be equally competitive, German and Netherlands research policy may need to respond to this.

At the end of this paper, we would like to address a point which was made by Sørensen and Wiborg Schneider (2017) and concerns all studies investigating the performance of countries. One can see it as a limitation of all these studies. The authors tried to find an answer on the question of how Danish research is. They show – for this small country – that "Danish research" is in 6 out of 10 cases research affiliated with at least one other country. Furthermore, close to 40% of new recruitments in Denmark are researchers with foreign citizenship. The authors conclude that "contemporary research has become a highly transnational activity" (Sørensen & Wiborg Schneider, 2017, p. 140). In other words, a paper assigned to a specific country based on the authors' affiliations is as a rule affiliated with other countries in any sort. It is thus increasingly difficult in bibliometric analysis to separate clear country effects. In this study, we included the number of countries mentioned on a paper as independent variable and thus considered that most of the papers are more or less "international".

Our study is based on papers published in the natural sciences in three citing years and references from three cited years (in order to focus on the research front). In future studies, it would be interesting to investigate whether the results which we found hold true for other citing years and especially other disciplines than natural sciences (such as medical and health sciences) and sub-disciplines. The OECD major codes which we used in this study to separate natural sciences papers are field categorizations on a very high level. Future studies could focus on the OECD minor codes which differentiate, e.g., natural sciences into mathematics, computer and information sciences, physical sciences and astronomy, chemical sciences, earth and related environmental sciences, and biological sciences. With regard to the cited references, it would be interesting to include in future studies not only the cited references from the most recent years, but also from a longer time ago.

# Acknowledgements

The bibliometric data used in this paper are from an in-house database developed and maintained by the Max Planck Digital Library (MPDL, Munich) and derived from the Science Citation Index Expanded (SCI-E), Social Sciences Citation Index (SSCI), Arts and Humanities Citation Index (AHCI) prepared by Clarivate Analytics, formerly the IP & Science business of Thomson Reuters (Philadelphia, Pennsylvania, USA).

# 6 Appendix

## 6.1 Key values

Table 7. Key values for the dependent and independent variables included in the regression analysis for Germany. In case of binary variables, the mean can be interpreted as a percentage.

| Variable | Mean | Standard deviation | Minimum | Maximum |
|---|---|---|---|---|
| Dependent variable | | | | |
| Highly-cited (citing) paper | 0.03 | 0.17 | 0 | 1 |
| Independent variables | | | | |
| Citing paper level | | | | |
| Publication year | | | | |
| 2009 | 0.31 | 0.46 | 0 | 1 |
| 2014 | 0.46 | 0.50 | 0 | 1 |
| Number of countries | 2.41 | 2.74 | 1 | 46 |
| Cited reference level | | | | |
| Country | | | | |
| USA | 0.37 | 0.48 | 0 | 1 |
| Germany | 0.37 | 0.48 | 0 | 1 |
| UK | 0.13 | 0.34 | 0 | 1 |
| France | 0.10 | 0.30 | 0 | 1 |
| Japan | 0.07 | 0.26 | 0 | 1 |
| China | 0.06 | 0.24 | 0 | 1 |
| Italy | 0.06 | 0.25 | 0 | 1 |
| Canada | 0.05 | 0.22 | 0 | 1 |
| Switzerland | 0.06 | 0.23 | 0 | 1 |
| The Netherlands | 0.05 | 0.22 | 0 | 1 |
| Years back of cited reference year | | | | |
| 2 | 0.36 | 0.48 | 0 | 1 |
| 3 | 0.33 | 0.47 | 0 | 1 |
| | | | | |
| Number of countries | 1.86 | 2.24 | 1 | 62 |
| English paper | 1.00 | 0.05 | 0 | 1 |
| Citation percentile | 80.29 | 19.76 | 2.11 | 100 |

Table 8. Key figures for the dependent and independent variables included in the regression analysis for the Netherlands. In case of binary variables, the mean can be interpreted as a percentage.

| Variable | Mean | Standard deviation | Minimum | Maximum |
|---|---|---|---|---|
| Dependent variable | | | | |
| Highly-cited (citing) paper | 0.05 | 0.21 | 0 | 1 |
| Independent variables | | | | |
| Citing paper level | | | | |
| Publication year | | | | |
| 2009 | 0.30 | 0.46 | 0 | 1 |
| 2014 | 0.48 | 0.50 | 0 | 1 |
| Number of countries | 3.08 | 3.93 | 1 | 46 |
| Cited reference level | | | | |
| Country | | | | |
| USA | 0.41 | 0.49 | 0 | 1 |
| The Netherlands | 0.26 | 0.44 | 0 | 1 |
| Germany | 0.18 | 0.38 | 0 | 1 |
| UK | 0.17 | 0.37 | 0 | 1 |
| France | 0.11 | 0.31 | 0 | 1 |
| Japan | 0.06 | 0.24 | 0 | 1 |
| China | 0.06 | 0.23 | 0 | 1 |
| Italy | 0.07 | 0.26 | 0 | 1 |
| Canada | 0.06 | 0.25 | 0 | 1 |
| Spain | 0.06 | 0.24 | 0 | 1 |
| Years back of cited reference year | | | | |
| 2 | 0.36 | 0.48 | 0 | 1 |
| 3 | 0.34 | 0.47 | 0 | 1 |
| | | | | |
| Number of countries | 2.04 | 2.64 | 1 | 52 |
| English paper | 1.00 | 0.03 | 0 | 1 |
| Citation percentile | 81.32 | 19.40 | 2.87 | 100 |

Table 9. Key figures for the dependent and independent variables included in the regression analysis for UK. In case of binary variables, the mean can be interpreted as a percentage.

| Variable | Mean | Standard deviation | Minimum | Maximum |
| --- | --- | --- | --- | --- |
| Dependent variable | | | | |
| Highly-cited (citing) paper | 0.04 | 0.19 | 0 | 1 |
| Independent variables | | | | |
| Citing paper level | | | | |
| Publication year | | | | |
| 2009 | 0.31 | 0.46 | 0 | 1 |
| 2014 | 0.45 | 0.50 | 0 | 1 |
| Number of countries | 2.53 | 2.89 | 1 | 46 |
| Cited reference level | | | | |
| Country | | | | |
| USA | 0.41 | 0.49 | 0 | 1 |
| UK | 0.34 | 0.48 | 0 | 1 |
| Germany | 0.14 | 0.35 | 0 | 1 |
| France | 0.10 | 0.30 | 0 | 1 |
| Japan | 0.06 | 0.24 | 0 | 1 |
| China | 0.06 | 0.24 | 0 | 1 |
| Canada | 0.07 | 0.25 | 0 | 1 |
| Italy | 0.07 | 0.25 | 0 | 1 |
| Spain | 0.06 | 0.23 | 0 | 1 |
| Australia | 0.05 | 0.22 | 0 | 1 |
| Years back of cited reference year | | | | |
| 2 | 0.36 | 0.48 | 0 | 1 |
| 3 | 0.34 | 0.47 | 0 | 1 |
| | | | | |
| Number of countries | 1.90 | 2.26 | 1 | 52 |
| English paper | 1.00 | 0.03 | 0 | 1 |
| Citation percentile | 80.75 | 19.63 | 2.87 | 100 |

## 6.2 Robustness checks

Table 10. Regression models for Germany including five (most frequently referenced), 15 (five randomly selected), and 20 (ten randomly selected) countries

|  | Five countries | | 15 countries | | 20 countries | |
|---|---|---|---|---|---|---|
|  | Odds ratio, 95% CI | Odds ratio, 95% CI | Odds ratio, 95% CI | Odds ratio, 95% CI | Odds ratio, 95% CI | Odds ratio, 95% CI |
|  | Excluding paper percentile | Including paper percentile | Excluding paper percentile | Including paper percentile | Excluding paper percentile | Including paper percentile |
| Citing paper level | | | | | | |
| Publication year | | | | | | |
| 2009 | 1.16 | 1.15 | 1.15 | 1.14 | 1.15 | 1.14 |
|  | [0.99,1.37] | [0.98,1.35] | [0.98,1.36] | [0.97,1.34] | [0.98,1.35] | [0.97,1.34] |
| 2014 | 1.50*** | 1.44*** | 1.46*** | 1.41*** | 1.45*** | 1.41*** |
|  | [1.27,1.76] | [1.23,1.70] | [1.24,1.71] | [1.20,1.66] | [1.24,1.71] | [1.20,1.66] |
| Number of countries | 1.12*** | 1.12*** | 1.12*** | 1.12*** | 1.12*** | 1.12*** |
|  | [1.10,1.14] | [1.09,1.14] | [1.10,1.14] | [1.10,1.14] | [1.10,1.14] | [1.10,1.14] |
| Cited reference level | | | | | | |
| Country | | | | | | |
| USA | 1.49*** | 1.23*** | 1.54*** | 1.26*** | 1.57*** | 1.27*** |
|  | [1.39,1.60] | [1.14,1.32] | [1.44,1.65] | [1.17,1.35] | [1.46,1.68] | [1.19,1.37] |
| Germany | 0.87*** | 0.87*** | 0.90** | 0.90*** | 0.92** | 0.91** |
|  | [0.81,0.92] | [0.82,0.93] | [0.85,0.96] | [0.85,0.95] | [0.87,0.98] | [0.86,0.97] |
| UK | 1.37*** | 1.24*** | 1.38*** | 1.24*** | 1.39*** | 1.25*** |
|  | [1.28,1.48] | [1.15,1.33] | [1.28,1.48] | [1.15,1.33] | [1.30,1.49] | [1.16,1.34] |
| France | 1.16*** | 1.11* | 1.18*** | 1.12** | 1.20*** | 1.13** |
|  | [1.07,1.26] | [1.02,1.20] | [1.09,1.28] | [1.03,1.21] | [1.11,1.29] | [1.05,1.22] |
| Japan | 1.08 | 1.04 | 1.12* | 1.06 | 1.14** | 1.08 |

|  | [1.00,1.18] | [0.96,1.13] | [1.03,1.22] | [0.98,1.16] | [1.04,1.24] | [0.99,1.18] |
|---|---|---|---|---|---|---|
| China |  |  | 1.26*** | 1.19** | 1.29*** | 1.21** |
|  |  |  | [1.10,1.44] | [1.05,1.36] | [1.13,1.47] | [1.06,1.39] |
| Italy |  |  | 1.19*** | 1.16** | 1.21*** | 1.18*** |
|  |  |  | [1.08,1.30] | [1.06,1.28] | [1.10,1.32] | [1.08,1.29] |
| Canada |  |  | 1.32*** | 1.25*** | 1.34*** | 1.27*** |
|  |  |  | [1.21,1.45] | [1.14,1.37] | [1.22,1.47] | [1.15,1.39] |
| Switzerland |  |  | 1.05 | 0.93 | 1.07 | 0.94 |
|  |  |  | [0.95,1.16] | [0.83,1.03] | [0.97,1.18] | [0.85,1.04] |
| The Netherlands |  |  | 1.21*** | 1.10* | 1.22*** | 1.12* |
|  |  |  | [1.10,1.34] | [1.00,1.22] | [1.10,1.36] | [1.01,1.24] |
| Finland |  |  | 1.27** | 1.22* | 1.27** | 1.23* |
|  |  |  | [1.06,1.53] | [1.02,1.47] | [1.06,1.53] | [1.02,1.48] |
| Australia |  |  | 1.30*** | 1.21** | 1.31*** | 1.21** |
|  |  |  | [1.15,1.46] | [1.07,1.36] | [1.16,1.47] | [1.08,1.37] |
| Ireland |  |  | 1.07 | 1.09 | 1.09 | 1.11 |
|  |  |  | [0.87,1.31] | [0.89,1.33] | [0.89,1.34] | [0.90,1.35] |
| Singapore |  |  | 1.44*** | 1.26* | 1.45*** | 1.27* |
|  |  |  | [1.20,1.73] | [1.05,1.51] | [1.21,1.74] | [1.06,1.52] |
| Chile |  |  | 1.17 | 1.20 | 1.18 | 1.21 |
|  |  |  | [0.95,1.45] | [0.97,1.49] | [0.95,1.46] | [0.98,1.49] |
| Sweden |  |  |  |  | 1.14* | 1.08 |
|  |  |  |  |  | [1.01,1.28] | [0.95,1.21] |
| South Africa |  |  |  |  | 1.00 | 1.02 |
|  |  |  |  |  | [0.81,1.24] | [0.83,1.27] |
| New Zealand |  |  |  |  | 1.31* | 1.25 |
|  |  |  |  |  | [1.03,1.66] | [0.98,1.59] |
| Norway |  |  |  |  | 1.12 | 1.12 |
|  |  |  |  |  | [0.95,1.33] | [0.95,1.32] |
| Portugal |  |  |  |  | 1.23* | 1.19 |
|  |  |  |  |  | [1.02,1.48] | [0.98,1.44] |

| Years back of cited reference year | | | | | | |
|---|---|---|---|---|---|---|
| 2 | 0.91*** | 0.91*** | 0.91*** | 0.91*** | 0.91*** | 0.91*** |
|  | [0.87,0.94] | [0.87,0.94] | [0.88,0.95] | [0.88,0.95] | [0.88,0.95] | [0.88,0.95] |
| 3 | 0.78*** | 0.76*** | 0.78*** | 0.76*** | 0.78*** | 0.76*** |
|  | [0.75,0.82] | [0.73,0.80] | [0.75,0.82] | [0.73,0.80] | [0.75,0.82] | [0.73,0.80] |
| Number of countries | 0.94*** | 0.94*** | 0.91*** | 0.92*** | 0.89*** | 0.90*** |
|  | [0.93,0.96] | [0.92,0.96] | [0.88,0.93] | [0.89,0.94] | [0.87,0.92] | [0.88,0.93] |
| English paper | 2.19* | 0.85 | 2.16* | 0.86 | 2.16* | 0.86 |
|  | [1.09,4.38] | [0.43,1.68] | [1.08,4.33] | [0.44,1.69] | [1.08,4.32] | [0.44,1.69] |
| Citation percentile |  | 1.04*** |  | 1.04*** |  | 1.04*** |
|  |  | [1.03,1.04] |  | [1.03,1.04] |  | [1.03,1.04] |
| Observations | 973,296 | 973,296 | 973,296 | 973,296 | 973,296 | 973,296 |

Notes.

Exponentiated coefficients; 95% confidence intervals in brackets.

* $p < 0.05$, ** $p < 0.01$, *** $p < 0.001$

Table 11. Regression models for the Netherlands including 5 (most frequently referenced), 15 (5 randomly selected), and 20 (10 randomly selected) countries.

|  | Five countries | | 15 countries | | 20 countries | |
| --- | --- | --- | --- | --- | --- | --- |
|  | Odds ratio, 95% CI | Odds ratio, 95% CI | Odds ratio, 95% CI | Odds ratio, 95% CI | Odds ratio, 95% CI | Odds ratio, 95% CI |
|  | Excluding paper percentile | Including paper percentile | Excluding paper percentile | Including paper percentile | Excluding paper percentile | Including paper percentile |
| Citing paper level | | | | | | |
| Publication year | | | | | | |
| 2009 | 1.42** | 1.40** | 1.41** | 1.39** | 1.41** | 1.39** |
|  | [1.11,1.81] | [1.10,1.78] | [1.11,1.80] | [1.09,1.78] | [1.11,1.80] | [1.09,1.78] |
| 2014 | 1.84*** | 1.80*** | 1.79*** | 1.75*** | 1.78*** | 1.74*** |
|  | [1.43,2.37] | [1.39,2.32] | [1.39,2.30] | [1.36,2.26] | [1.39,2.29] | [1.36,2.24] |
| Number of countries | 1.10*** | 1.10*** | 1.11*** | 1.10*** | 1.11*** | 1.10*** |
|  | [1.08,1.13] | [1.08,1.13] | [1.08,1.13] | [1.08,1.13] | [1.08,1.13] | [1.08,1.13] |
| Cited reference level | | | | | | |
| Country | | | | | | |
| USA | 1.44*** | 1.24*** | 1.49*** | 1.28*** | 1.55*** | 1.33*** |
|  | [1.29,1.61] | [1.12,1.39] | [1.34,1.66] | [1.16,1.43] | [1.39,1.73] | [1.20,1.48] |
| The Netherlands | 0.84** | 0.84** | 0.87** | 0.87** | 0.90 | 0.90 |
|  | [0.75,0.93] | [0.76,0.94] | [0.78,0.96] | [0.78,0.97] | [0.80,1.00] | [0.80,1.00] |
| Germany | 1.06 | 1.00 | 1.09 | 1.02 | 1.12* | 1.05 |
|  | [0.95,1.19] | [0.89,1.12] | [0.97,1.21] | [0.92,1.14] | [1.00,1.25] | [0.95,1.18] |
| UK | 1.29*** | 1.20*** | 1.29*** | 1.20*** | 1.33*** | 1.24*** |
|  | [1.16,1.44] | [1.08,1.34] | [1.16,1.42] | [1.08,1.33] | [1.20,1.48] | [1.11,1.37] |
| France | 1.08 | 1.06 | 1.10 | 1.08 | 1.14* | 1.11 |
|  | [0.96,1.23] | [0.93,1.21] | [0.98,1.25] | [0.95,1.21] | [1.01,1.29] | [0.98,1.26] |
| Japan | | | 1.07 | 1.05 | 1.11 | 1.09 |
|  | | | [0.94,1.21] | [0.92,1.18] | [0.98,1.26] | [0.96,1.23] |
| China | | | 1.21 | 1.21 | 1.28 | 1.27 |
| | | | [0.92,1.59] | [0.92,1.59] | [0.98,1.67] | [0.97,1.66] |
| Italy | | | 1.21** | 1.22** | 1.25*** | 1.25*** |
| | | | [1.06,1.38] | [1.07,1.38] | [1.09,1.42] | [1.10,1.42] |
| Canada | | | 1.31*** | 1.27** | 1.34*** | 1.29*** |
| | | | [1.14,1.52] | [1.10,1.47] | [1.16,1.55] | [1.12,1.49] |
| Spain | | | 1.10 | 1.13 | 1.13 | 1.15 |
| | | | [0.96,1.27] | [0.98,1.31] | [0.98,1.30] | [1.00,1.33] |
| Ireland | | | 1.14 | 1.15 | 1.21 | 1.20 |
| | | | [0.85,1.54] | [0.86,1.55] | [0.90,1.61] | [0.90,1.60] |
| New Zealand | | | 1.37 | 1.31 | 1.38 | 1.32 |
| | | | [0.94,1.99] | [0.90,1.90] | [1.00,1.91] | [0.96,1.82] |
| Denmark | | | 1.67*** | 1.55** | 1.64*** | 1.52** |
| | | | [1.27,2.18] | [1.18,2.02] | [1.24,2.18] | [1.14,2.02] |
| Australia | | | 1.26** | 1.21* | 1.29** | 1.24** |
| | | | [1.08,1.48] | [1.04,1.42] | [1.09,1.52] | [1.05,1.46] |
| Singapore | | | 1.11 | 1.02 | 1.09 | 1.01 |
| | | | [0.83,1.48] | [0.77,1.36] | [0.80,1.49] | [0.74,1.36] |
| Norway | | | | | 1.39** | 1.41** |
| | | | | | [1.09,1.78] | [1.10,1.80] |
| Chile | | | | | 1.28 | 1.32 |
| | | | | | [0.92,1.78] | [0.95,1.82] |
| Portugal | | | | | 1.21 | 1.21 |
| | | | | | [0.90,1.61] | [0.90,1.62] |
| Austria | | | | | 1.52*** | 1.44*** |
| | | | | | [1.25,1.85] | [1.18,1.76] |
| Thailand | | | | | 2.27* | 2.26** |
| | | | | | [1.19,4.30] | [1.22,4.21] |
| Years back of cited reference year | | | | | | |
| 2 | 0.92** | 0.92** | 0.92** | 0.92** | 0.92** | 0.92** |
|                    | [0.86,0.97]    | [0.87,0.98]    | [0.86,0.97]    | [0.87,0.98]    | [0.86,0.97]    | [0.87,0.98]    |
| 3                  | 0.81***        | 0.79***        | 0.80***        | 0.79***        | 0.80***        | 0.79***        |
|                    | [0.75,0.87]    | [0.73,0.86]    | [0.74,0.87]    | [0.73,0.85]    | [0.74,0.86]    | [0.73,0.85]    |
| Number of countries| 0.94***        | 0.94***        | 0.90***        | 0.90***        | 0.86***        | 0.87***        |
|                    | [0.92,0.96]    | [0.92,0.96]    | [0.87,0.93]    | [0.87,0.93]    | [0.83,0.91]    | [0.83,0.91]    |
| English paper      | 4.32           | 2.04           | 4.20           | 2.01           | 4.30           | 2.07           |
|                    | [0.82,22.83]   | [0.38,11.07]   | [0.80,22.09]   | [0.37,10.82]   | [0.77,23.90]   | [0.36,11.92]   |
|                    |                |                |                |                |                |                |
| Citation percentile|                | 1.03***        |                | 1.03***        |                | 1.03***        |
|                    |                | [1.02,1.03]    |                | [1.02,1.03]    |                | [1.02,1.03]    |
| Observations       | 266,000        | 266,000        | 266,000        | 266,000        | 266,000        | 266,000        |

Notes.

Exponentiated coefficients; 95% confidence intervals in brackets.

* $p < 0.05$, ** $p < 0.01$, *** $p < 0.001$

Table 12. Regression models for the UK including 5 (most frequently referenced), 15 (5 randomly selected), and 20 (10 randomly selected) countries.

|  | Five countries | | 15 countries | | 20 countries | |
| --- | --- | --- | --- | --- | --- | --- |
|  | Odds ratio, 95% CI | Odds ratio, 95% CI | Odds ratio, 95% CI | Odds ratio, 95% CI | Odds ratio, 95% CI | Odds ratio, 95% CI |
|  | Excluding paper percentile | Including paper percentile | Excluding paper percentile | Including paper percentile | Excluding paper percentile | Including paper percentile |
| Citing paper level | | | | | | |
| Publication year | | | | | | |
| 2009 | 1.23** | 1.21* | 1.21* | 1.20* | 1.21* | 1.20* |
|  | [1.06,1.43] | [1.04,1.41] | [1.05,1.41] | [1.03,1.39] | [1.04,1.41] | [1.03,1.39] |
| 2014 | 1.56*** | 1.50*** | 1.51*** | 1.46*** | 1.50*** | 1.45*** |
|  | [1.35,1.81] | [1.29,1.74] | [1.31,1.75] | [1.26,1.69] | [1.30,1.74] | [1.25,1.68] |
| Number of countries | 1.11*** | 1.10*** | 1.11*** | 1.11*** | 1.11*** | 1.11*** |
|  | [1.09,1.13] | [1.09,1.12] | [1.09,1.13] | [1.09,1.13] | [1.09,1.13] | [1.09,1.13] |
| Cited reference level | | | | | | |
| USA | 1.53*** | 1.27*** | 1.61*** | 1.33*** | 1.62*** | 1.34*** |
|  | [1.43,1.64] | [1.19,1.36] | [1.51,1.71] | [1.25,1.42] | [1.52,1.73] | [1.26,1.44] |
| UK | 1.01 | 0.99 | 1.06* | 1.03 | 1.07* | 1.04 |
|  | [0.95,1.07] | [0.94,1.05] | [1.00,1.12] | [0.98,1.09] | [1.01,1.13] | [0.98,1.10] |
| Germany | 1.12** | 1.03 | 1.17*** | 1.06 | 1.18*** | 1.07* |
|  | [1.04,1.20] | [0.96,1.10] | [1.09,1.25] | [0.99,1.14] | [1.10,1.26] | [1.00,1.15] |
| France | 1.07 | 1.02 | 1.11** | 1.06 | 1.12** | 1.07 |
|  | [0.99,1.15] | [0.94,1.10] | [1.03,1.20] | [0.98,1.14] | [1.04,1.21] | [0.99,1.15] |
| Japan | 1.13** | 1.11** | 1.18*** | 1.15*** | 1.19*** | 1.16*** |
|  | [1.05,1.23] | [1.03,1.20] | [1.09,1.28] | [1.06,1.25] | [1.10,1.29] | [1.07,1.26] |
| China | | | 1.35*** | 1.30*** | 1.35*** | 1.30*** |
|  | | | [1.18,1.54] | [1.13,1.48] | [1.18,1.55] | [1.14,1.49] |
| Canada | | | 1.31*** | 1.24*** | 1.32*** | 1.24*** |
| | | | [1.21,1.42] | [1.14,1.34] | [1.21,1.43] | [1.15,1.35] |
| Italy | | | 1.03 | 1.02 | 1.05 | 1.03 |
| | | | [0.95,1.13] | [0.94,1.11] | [0.96,1.14] | [0.95,1.13] |
| Spain | | | 1.18** | 1.18*** | 1.18** | 1.19*** |
| | | | [1.07,1.30] | [1.07,1.31] | [1.07,1.30] | [1.08,1.31] |
| Australia | | | 1.44*** | 1.34*** | 1.44*** | 1.35*** |
| | | | [1.31,1.59] | [1.22,1.48] | [1.31,1.59] | [1.22,1.49] |
| Finland | | | 1.32*** | 1.29** | 1.32*** | 1.29** |
| | | | [1.13,1.55] | [1.10,1.51] | [1.12,1.54] | [1.10,1.51] |
| Thailand | | | 1.42 | 1.40 | 1.38 | 1.39 |
| | | | [0.97,2.08] | [0.96,2.06] | [0.94,2.04] | [0.95,2.04] |
| Austria | | | 1.11 | 1.06 | 1.14 | 1.09 |
| | | | [0.96,1.29] | [0.91,1.22] | [0.99,1.32] | [0.94,1.25] |
| Chile | | | 1.21 | 1.26* | 1.22* | 1.27* |
| | | | [0.99,1.48] | [1.03,1.54] | [1.01,1.49] | [1.04,1.54] |
| Norway | | | 1.23** | 1.21* | 1.23** | 1.22* |
| | | | [1.05,1.44] | [1.04,1.42] | [1.05,1.44] | [1.04,1.42] |
| Ireland | | | | | 1.23* | 1.23* |
| | | | | | [1.01,1.49] | [1.02,1.48] |
| Singapore | | | | | 1.75*** | 1.53*** |
| | | | | | [1.41,2.16] | [1.23,1.90] |
| South Africa | | | | | 1.05 | 1.07 |
| | | | | | [0.87,1.27] | [0.89,1.30] |
| Portugal | | | | | 1.08 | 1.10 |
| | | | | | [0.90,1.30] | [0.91,1.32] |
| New Zealand | | | | | 1.06 | 1.04 |
| | | | | | [0.90,1.25] | [0.88,1.22] |
| Years back of cited reference year | . | . | . | . | | |
| 2 | 0.92*** | 0.91*** | 0.92*** | 0.92*** | 0.92*** | 0.92*** |
| | [0.88,0.95] | [0.88,0.95] | [0.89,0.95] | [0.88,0.95] | [0.89,0.95] | [0.88,0.95] |
| 3 | 0.78*** | 0.76*** | 0.78*** | 0.76*** | 0.78*** | 0.76*** |
| | [0.74,0.81] | [0.72,0.80] | [0.74,0.82] | [0.72,0.80] | [0.74,0.82] | [0.72,0.80] |
| Number of countries | 0.95*** | 0.95*** | 0.90*** | 0.90*** | 0.89*** | 0.90*** |
| | [0.93,0.96] | [0.93,0.96] | [0.88,0.92] | [0.88,0.93] | [0.87,0.92] | [0.87,0.92] |
| English paper | 1.39 | 0.59 | 1.42 | 0.61 | 1.41 | 0.61 |
| | [0.57,3.34] | [0.25,1.39] | [0.59,3.43] | [0.26,1.44] | [0.58,3.41] | [0.26,1.44] |
| | | | | | | |
| Citation percentile | | 1.04*** | | 1.04*** | | 1.04*** |
| | | [1.03,1.04] | | [1.03,1.04] | | [1.03,1.04] |
| Observations | 840,697 | 840,697 | 840,697 | 840,697 | 840,697 | 840,697 |

Notes.
Exponentiated coefficients; 95% confidence intervals in brackets.
$^{*} p < 0.05$, $^{**} p < 0.01$, $^{***} p < 0.001$